\documentclass[aps,prd,nofootinbib,floatfix,amsmath,amssymb,colorBG,spanish]{revtex4}
\usepackage{amssymb}
\usepackage{graphicx}
\usepackage{color}
\usepackage[tight,TABTOPCAP]{subfigure}
\usepackage{epstopdf}
\begin{document}

\title{Heavy neutral scalar decays into electroweak gauge bosons in the Littlest Higgs Model}

\author{$^{(a)}$J. I. Aranda, $^{(a,b)}$I. Cort\'es-Maldonado,$^{(a)}$S. Montejo-Montejo,
$^{(a)}$F. Ram\'irez-Zavaleta, $^{(a)}$E. S. Tututi}

\address{$^{(a)}$Facultad de Ciencias F\'isico Matem\'aticas, Universidad Michoacana de San Nicol\'as de Hidalgo,
Avenida Francisco J. M\'ujica S/N, 58060, Morelia, Michoac\'an, M\'exico.\\
$^{(b)}$C\'atedras Consejo Nacional de Ciencia y Tecnolog\'ia, M\'exico.}

\date{\today}
\begin{abstract}
We study the heavy neutral scalar decays into standard model electroweak gauge bosons in the context of the Littlest
Higgs model. We focus our attention on the $\Phi^0 \to WW, \gamma V$ processes induced at the one-loop level, with $V=\gamma, Z$. Since the branching ratios of the $\Phi^0 \to \gamma V$ decays result very suppressed, only the $\Phi^0 \to WW$ process is analyzed in the framework of possible experimental scenarios by using heavy scalar masses between 1.6 TeV until 3.3 TeV. The branching ratio for the $\Phi^0 \to WW$ decay is of the order of $10^{-3}$ throughout the interval 2 TeV $< f <$ 4 TeV, which represent the global symmetry breaking scale of the theory. Thus, it is estimated the associated production cross section for the $pp\to \Phi^0 X\to WW$, finding around ten events for $m_{\Phi^0}\approx 1.6$ TeV at best.

\end{abstract}


\maketitle
\section{Introduction}
At the LHC, the work of many experimental collaborations that have been dedicated their efforts to the search for the Higgs boson was succeeded~\cite{Aad:2012tfa,Chatrchyan:2012xdj}. This particle represents the first evidence of detection of a spin-0 particle, which in principle seems to be the Higgs boson of the SM. In this direction, one could think that in the fundamental theory (at high energies), there may exist couplings between new scalar particles with all the particles detected. In addition, since the LHC is the first machine that operates at the scale of TeVs, it is hoped to have good possibilities for detection of new particles at this energy scale. Therefore, it is expected that in the following years the LHC can offer different possibilities of detection of this type of particles or discard their existence at the scale of TeVs. At least, the scenario is optimistic since the experimental collaborations ATLAS and CMS have been able to construct a reliable detection machinery referring to spin-0 resonances. Indeed, ATLAS and CMS collaborations have continued searching for new exotic particles, such as the Randall-Sundrum spin-2 boson~\cite{CMS:2015dxe} or new heavy scalar particles~\cite{atlas1}. Thus, the possibility to search for new physics phenomena at this energy scale is higher than past years with the upgrade of the CMS and ATLAS detectors~\cite{ATLAS:2013hta,Khachatryan:2014hpa,CMS:2014wda,Aad:2015owa}. Abreast, in the literature there are theoretical proposals for the search of new heavy scalar particles~\cite{SMHS}. In more recent works, with the aim of the explaining a possible resonance around 750 GeV in the diphoton channel, which was misunderstood due to statistical fluctuation, a lot of phenomenological studies were done~\cite{Low:2015qho,Ben-Dayan,Qing,Fichet:2016pvq,Cao:2016udb,Ding:2016ldt,Davis:2016hlw,Ren:2016gyg}.

Several extensions of the standard model (SM) predict the existence of new particles with masses of the order of TeVs~\cite{2HDM,LR1,LR2,MSSM1,Instantons,MSSM2,ZP,331M,3311M}. The exhaustive search for new scalar particles at the LHC give us the possibility to explore new physics processes related to extended Higgs sectors. In this sense, there are many alternative formulations that predict more content of scalar particles such as the two-Higgs doublet models (2HDMs)~\cite{2HDM, 2HDMs1}, three-Higgs doublet model (3HDM)~\cite{3HDMs}, Higgs-singlet extension model~\cite{Pruna}, little Higgs models (LHMs)~\cite{LHMs}, etc. Among the wide variety of LHMs, the case of the littlest Higgs model (LTHM) is of peculiar interest since there is no new degrees of freedom beyond the SM under TeV scale together with its reduced spectrum of new scalar particles~\cite{LHM1}. Furthermore, this type of models offer a possible solution to hierarchy problem, which takes place when the Higgs boson mass is affected by corrections at one-loop level.

The LHMs also arise as an alternative for the study of the electroweak symmetry breaking \cite{LHMs,LHM1}, on the basis of dimensional deconstruction~\cite{Dec1,Dec2}, with feature of canceling quadratic divergences. In fact, the quadratic divergence generated at the one-loop level through the SM gauge bosons is canceled via the quadratic divergence introduced by the new heavy gauge bosons at the same perturbative level. The one-loop quadratic divergence induced in the SM Yukawa sector is removed by introducing new heavy fermions such that it cancels the quadratic divergence coming from the top quark. In LHMs, the new Higgs field acquire mass becoming pseudo-Goldstone bosons in accordance with the breaking of a global symmetry at the energy scale of TeVs, where a massless Higgs appears. The quadratic-divergent corrections to the Higgs mass arise at loop level, therefore, this warrants a light Higgs. With regard to LTHM~\cite{Han:2003wu}, the new states arising at the TeV energy scale form a new set of four gauge bosons with the same quantum numbers as the SM gauge bosons, namely, $A_H$, $Z_H$, and $W^{\pm}_H$, an exotic quark with the same charge as the SM top quark, and a new heavy scalar triplet, which contains six physical states: double-charged scalars $\Phi^{\pm\pm}$, single-charged scalars $\Phi^\pm$, a neutral scalar $\Phi^0$, and a neutral pseudo scalar $\Phi^p$. A detailed presentation of LTHM model can be found in Ref.~\cite{Han:2003wu}.

The main objective of this manuscript is to study the relevance of the $\Phi^0\to WW, \gamma V$ ($V=\gamma, Z$) decays in the context of the linearized theory of the LTHM, which refers to make a first-order expansion of the $\Sigma$ field around its vacuum expectation in powers of $v/f$~\cite{Han:2003wu}. Currently, the parameter space of the LTHM has been severely constrained by the Higgs discovery channels and electroweak precision observables~\cite{Reuter}, therefore, the processes in question serve as a way to test this model. Our analysis considers pooled results from experimental and phenomenological studies where an experimental scenario as realistic as possible is set~\cite{Reuter}. The viable interval of experimental analysis for the energy scale $f$ comes from a phenomenological study based on the experimental searches of a scalar boson consistent with SM-Higgs boson by making use of the signal strength modifier along with electroweak precision data. In order to be consistent with electroweak precision data, a lower limit for the energy scale $f$ around 2-4 TeV is established~\cite{Reuter}. As an added value, we present a phenomenological scenario on the $WW$ associated production at LHC resulting from the hypothetical $\Phi^0$ resonance with $pp$ collisions at center of mass energy of 14 TeV. Thereby, the phenomenological implications of a heavy neutral scalar boson with mass around 1 TeV decaying into two $W$ bosons are discussed in the LHC setting.

The present paper is organized as follows. In section~\ref{MOD-FRA}, we present an overview of the littlest Higgs model with special attention on the scalar sector. In section~\ref{ANA-CAL}, we analyze the $\Phi^0 \to WW, \gamma V$ decays at one-loop level. In section~\ref{NUM-RES}, the numerical results are discussed. Finally, the conclusions are presented in section~\ref{CONCLU}.


\section{Model framework}\label{MOD-FRA}
The LTHM is constructed on the basis of a nonlinear sigma model with $SU(5)$ global symmetry together with the gauged group $[SU(2)_1\otimes U(1)_1]\otimes [SU(2)_2\otimes U(1)_2]$~\cite{LHM1, Han:2003wu}. The $SU(5)$ group is spontaneously broken to $SO(5)$ at the energy scale $f$, where $f$ is constrained to be of the order of $2\textendash4$ TeV~\cite{Reuter}. At the same time, the $[SU(2)_1\otimes U(1)_1]\otimes [SU(2)_2\otimes U(1)_2]$ group is also broken to its subgroup $SU_L(2)\otimes U_Y(1)$, which is identified as the SM electroweak gauge group. The global symmetry breaking pattern leaves 14 Goldstone bosons which transform under the $SU_L(2)\otimes U_Y(1)$ group as a real singlet $\mathbf{1}_0$, a real triplet $\mathbf{3}_0$, a complex doublet $\mathbf{2}_{\pm \frac{1}{2}}$, and a complex triplet $\mathbf{3}_{\pm 1}$~\cite{LHM1,Han:2003wu}. At the scale $f$, the spontaneous global symmetry breaking of the $SU(5)$ group is generated by the vacuum expectation value (VEV) of the $\Sigma$ field, denoted as $\Sigma_0$~\cite{Han:2003wu}. The $\Sigma$ field is explicitly given by
\begin{equation}
\Sigma=e^{i\Pi/f}\Sigma_0e^{i\Pi^T/f},
\end{equation}
with
\begin{equation}
\Sigma_0=\left(\begin{array}{ccc}
\mathbf{0}_{2\times 2} & \mathbf{0}_{2\times 1} & \mathbf{1}_{2\times 2}\\
\mathbf{0}_{1\times 2} & 1 & \mathbf{0}_{1\times 2}\\
\mathbf{1}_{2\times 2} & \mathbf{0}_{2\times 1} & \mathbf{0}_{2\times 2}
\end{array}\right)
\end{equation}
and $\Pi$ being the Goldstone boson matrix having the following form
\begin{equation}
\Pi=\left(\begin{array}{ccc}
\mathbf{0}_{2\times 2} & h^\dagger/\sqrt{2} & \phi^\dagger\\
h/\sqrt{2} & 0 & h^\ast/\sqrt{2}\\
\phi & h^T/\sqrt{2} & \mathbf{0}_{2\times 2}
\end{array}\right).
\end{equation}
Here, $h$ is a doublet and $\phi$ represents a triplet under the $SU_L(2)\otimes U_Y(1)$ SM gauge group~\cite{Han:2003wu}
\begin{equation}
h = (h^+, h^0), \qquad
\phi = \left( \begin{array}{cc}
\phi^{++} & \frac{\phi^+}{\sqrt{2}} \\
\frac{\phi^+}{\sqrt{2}} & \phi^0
\end{array} \right).
\end{equation}
By the spontaneous symmetry breaking (SSB), both the real singlet and the
real triplet are absorbed by the longitudinal components of the gauge bosons at the energy scale $f$.
At this scale, the complex doublet and the complex triplet remain massless. The complex
triplet acquires a mass of the order of $f$ by means of the Coleman-Weinberg type potential when the
global symmetry of the group $SO(5)$ breaks down. The complex doublet is identified as the SM Higgs field.

The effective Lagrangian invariant under the $[SU(2)_1\otimes U(1)_1]\otimes [SU(2)_2\otimes U(1)_2]$
group is~\cite{Han:2003wu}
\begin{equation}
\mathcal{L}_{LTHM}=\mathcal{L}_G+\mathcal{L}_F+\mathcal{L}_\Sigma+\mathcal{L}_Y-V_{CW},
\end{equation}
where $\mathcal{L}_G$ represents the gauge bosons kinetic contributions, $\mathcal{L}_F$ the
fermion kinetic contributions, $\mathcal{L}_\Sigma$ the nonlinear sigma model contributions
of the LTHM, $\mathcal{L}_Y$ the Yukawa couplings of fermions and pseudo-Goldstone bosons,
and the last term symbolizes the Coleman-Weinberg potential.

The standard form of the Lagrangian of the nonlinear sigma model is
\begin{equation}
\mathcal{L}_\Sigma=\frac{f^2}{8}\mathrm{tr}\left|\mathcal{D}_\mu\Sigma\right|^2,
\end{equation}
where the covariant derivative is written as
\begin{equation}
\mathcal{D}_\mu\Sigma=\partial_\mu\Sigma-i\sum\limits_{j=1}^{2}\left[g_j \sum\limits_{a=1}^{3}W_{\mu j}^{a}\left(Q_j^a\Sigma+\Sigma Q_j^{a T}\right)+g^\prime_j B_{\mu j}\left(Y_j\Sigma+\Sigma Y_j^{T}\right)\right].
\end{equation}
Here, $W_{\mu j}^{a}$ are the $SU(2)$ gauge fields, $B_{\mu j}$ are the $U(1)$ gauge fields, $Q_j^a$ are the $SU(2)$ gauge group generators, $Y_j$ are the $U(1)$ gauge group generators, $g_j$ are the coupling constants of the $SU(2)$ group, and $g_j^\prime$ are the coupling constants of the $U(1)$ group~\cite{Han:2003wu}. After SSB around $\Sigma_0$, it is generated the mass eigenstates of order $f$ for the gauge bosons~\cite{Han:2003wu}
\begin{eqnarray}
W^\prime_\mu&=&-cW_{\mu 1}+sW_{\mu 2},\\
B^\prime_\mu&=&-c^\prime \mathcal{B}_{\mu 1}+s^\prime \mathcal{B}_{\mu 2},\\
W_\mu&=& sW_{\mu 1}+cW_{\mu 2},\\
B_\mu&=& s^\prime \mathcal{B}_{\mu 1}+c^\prime \mathcal{B}_{\mu 2},
\end{eqnarray}
where $W_{\mu j}\equiv\sum\limits_{a=1}^{3}W^a_{\mu j}Q^a_j$ and $\mathcal{B}_{\mu j}\equiv B_{\mu j}Y_j$ for $j=1,2$; $c=g_1/\sqrt{g_1^2+g_2^2}$, $c^\prime=g^\prime_1/\sqrt{g_1^{\prime 2}+g_2^{\prime 2}}$, $s=g_2/\sqrt{g_1^2+g_2^2}$, and $s^\prime=g^\prime_2/\sqrt{g_1^{\prime 2}+g_2^{\prime 2}}$. Notice that $\Sigma$ field has been expanded around $\Sigma_0$ holding dominant terms in $\mathcal{L}_\Sigma$~\cite{Han:2003wu}. At this stage of SSB the $B_\mu$ and $W_\mu$ fields remain massless.

The SSB at the Fermi scale provides mass to the SM gauge bosons ($B$ and $W$) and induces mixing
between heavy and light gauge bosons. The arising masses at the leading order (neglecting terms
of order $\mathcal{O}\left(\frac{v^2}{f^2}\right)$, with $v$ being the vacuum expectation value
at the Fermi scale) are 
\begin{eqnarray}
m_{Z_H}&=&\frac{gf}{2sc},\\
m_{A_H}&=&\frac{g^\prime f}{2\sqrt{5}s^\prime c^\prime},\\
m_{W_H}&=&\frac{gf}{2sc}. 
\end{eqnarray}
As it is known, $c=m_{W_{H}}/m_{Z_{H}}$ and takes the value equals to one at the leading order~\cite{Han:2003wu, Aranda}, in order to have values for the masses of the weak gauge bosons not very different, as it occurs in the electroweak sector of the SM~\cite{Aranda}.

In LTHM the Higgs potential is generated by one-loop radiative corrections at the leading order. This potential contains the contributions coming from gauge boson loops and fermion loops. When the $\Sigma$ field is expanded into the nonlinear sigma model it is obtained the associated Coleman-Weinberg potential~\cite{Coleman:1973jx}
\begin{equation}
 V_{CW}=\lambda_{\phi^2}f^2\mbox{Tr}(\phi^\dagger \phi)+
 i\lambda_{h\phi h}f(h\phi^\dagger h^T-h^* \phi^\dagger h^\dagger) -
 \mu^2hh^\dagger+\lambda_h^4(hh^\dagger)^2,
\end{equation}
wherein the $\lambda$'s coefficient are given by
\begin{eqnarray}
\lambda_{\phi^2} &=& \frac{a}{2} \left[ \frac{g^2}{s^2c^2}
+ \frac{g^{\prime 2}}{s^{\prime 2}c^{\prime 2}} \right]
+ 8 a^{\prime} \lambda_1^2,
\nonumber \\
\lambda_{h \phi h} &=& -\frac{a}{4}
\left[ g^2 \frac{(c^2-s^2)}{s^2c^2}
+ g^{\prime 2} \frac{(c^{\prime 2}-s^{\prime 2})}
{s^{\prime 2}c^{\prime 2}} \right]
+ 4 a^{\prime} \lambda_1^2,
\nonumber \\
\lambda_{h^4} &=& \frac{a}{8} \left[ \frac{g^2}{s^2c^2}
+ \frac{g^{\prime 2}}{s^{\prime 2}c^{\prime 2}} \right]
+ 2 a^{\prime} \lambda_1^2 = \frac{1}{4} \lambda_{\phi^2},
\end{eqnarray}
where $c$, $s$ ($c^\prime$, $s^\prime$) are the gauge coupling constants of the $SU(2)$ ($U(1)$) symmetry group. The $a$ and $a^\prime$ parameters represent the unknown ultraviolet (UV) physics at the cutoff scale $\Lambda_S$. Their values depend on the details of the UV completion at the cutoff scale $\Lambda_S$~\cite{Han:2003wu}. As far as the $\mu^2$ parameter refers, it is a free parameter which receives equally significant contributions from one-loop logarithmic and two-loop quadratically divergent parts~\cite{Han:2003wu}. The VEV $v$ ($v^\prime$) of the doublet (of the triplet) is obtained after minimizing the $V_{CW}$ potential, which fulfill the following relations
\begin{equation}
v^2=\frac{\mu^2}{\lambda_{h^4} -\frac{\lambda^2_{h\phi h}}{\lambda_{\phi^2}}},
\qquad v^\prime=\frac{\lambda_{h\phi h}v^2}{2\lambda_{\phi^2} f}.
\end{equation}
The masses of the heavy scalars are obtained by diagonalizing the Higgs mass matrix~\cite{Han:2003wu}. Thus, the gauge eigenstates of the Higgs sector can be written in terms of the mass eigenstates in the following way
\begin{eqnarray}
	h^0 &=& \frac{1}{\sqrt{2}}\left( c_0 H - s_0 \Phi^0 + v \right)
	+ \frac{i}{\sqrt{2}} \left( c_P G^0 - s_P \Phi^P \right), \nonumber \\
	\phi^0 &=& \frac{1}{\sqrt{2}}\left( s_P G^0 + c_P \Phi^P \right)
	- \frac{i}{\sqrt{2}} \left( s_0 H + c_0 \Phi^0 + \sqrt{2} v^{\prime} \right),
	\nonumber \\
	h^+ &=& c_+ G^+ - s_+ \Phi^+, \nonumber\\
	\phi^+ &=&\frac{1}{i}\left( s_+ G^+ + c_+ \Phi^+ \right),  \nonumber \\
	\phi^{++} &=& \frac{\Phi^{++}}{i},
\end{eqnarray}
where $H$ is the Higgs boson, $\Phi^0$ is a new neutral scalar, $\Phi^P$ is a neutral pseudoscalar, $\Phi^+$ and $\Phi^{++}$ are the charged and doubly charged scalars, and $G^+$ and $G^0$ are the Goldstone bosons that are eaten by the massless $W$ and $Z$ bosons~\cite{Han:2003wu}. At the leading order in the theory, the masses of the new heavy scalar particles are degenerate~\cite{Han:2003wu}
\begin{equation}
 m_\Phi=\frac{\sqrt{2} m_H}{\sqrt{{1-y_v^2}}} \frac{f}{v},
\label{massphi}
\end{equation}
where $y_v=4v^{\prime} f / v^2$. This mass expression is positive definite if
\begin{equation}
{v'^2\over v^2} < { v^2\over 16f^2}.
\end{equation}
Additionally, the generic expression for the Higgs mass is given by
\begin{equation}
m^2_{H} \simeq 2 \left( \lambda_{h^4}
- \lambda_{h \phi h}^2 / \lambda_{\phi^2} \right) v^2 = 2 \mu^2.
\end{equation}

Continuing with the description of the model, the LTHM incorporates new heavy fermions which couple to Higgs field in a such way that the quadratic
divergence of the top quark is canceled~\cite{LHM1,Han:2003wu}. In particular, this model introduces a
new set of heavy fermions arranged as a vector-like pair ($\tilde{t},\tilde{t}^{\prime c}$) with quantum
numbers $(\mathbf{3},\mathbf{1})_{Y_i}$ and $(\bar{\mathbf{3}},\mathbf{1})_{-Y_i}$, respectively. The new
Yukawa interactions are proposed to be
\begin{equation}
\mathcal{L}_Y=\frac{1}{2}\lambda_1\, f\,\epsilon_{ijk}\epsilon_{xy}\,\chi_i\,\Sigma_{jx}\,\Sigma_{ky}u_3^{\prime c}+\lambda_2\,f\,\tilde{t}\tilde{t}^{\prime c}+\mathrm{H.c.},
\end{equation}
where $\chi_i=(b_3,t_3,\tilde{t})$; $\epsilon_{ijk}$ and $\epsilon_{xy}$ are antisymmetric
tensors for $i,j,k=1,2,3$ and $x,y=4,5$~\cite{LHM1}. Here, $\lambda_1$ and $\lambda_2$ are free
parameters, where the $\lambda_2$ parameter can be fixed such that, for given $(f,\lambda_1)$, the
top quark mass adjust to its experimental value~\cite{Reuter}.

Expanding the $\Sigma$ field and retaining terms up to $\mathcal{O}(v^2/f^2)$ after diagonalizing the
mass matrix, it can be obtained the mass states $t_L$, $t^c_R$, $T_L$, and $T^c_R$, which correspond to
SM top quark and the heavy top quark, respectively~\cite{Han:2003wu,Reuter}. The explicit remaining terms of the Lagrangian $\mathcal{L}_{LTHM}$ as well as the complete set of new Feynman rules can be found in Ref.~\cite{Han:2003wu}.

\section{The study of the $\Phi^0\to WW, \gamma V$ decays}\label{ANA-CAL}
To study the one-loop level $\Phi^0\to WW, \gamma V$ decays we begin with the calculation of the total decay width of the $\Phi^0$ ($\Gamma_{\Phi^0}$) which includes only SM final states. The dominant contributions to $\Gamma_{\Phi^0}$ are the tree-level decays of $\Phi^0$ into $HH, ZZ, ZZH$. However, we will also compute the subdominant contributions of final states such as $\bar{t}t, WWZ, WWH$.

\subsubsection{Tree-level decays}
\begin{figure}[!ht]
\begin{center}
\includegraphics[scale=0.55]{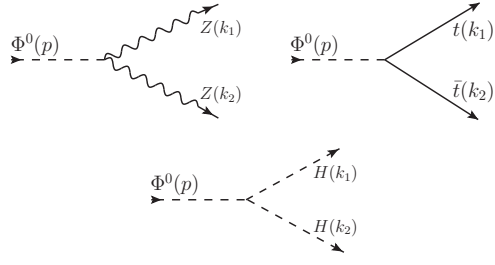}
\caption{Feynman diagrams contributing to the $\Phi^0 \to t \bar t, ZZ, HH$ decays at tree level.}
\label{feyn2VV}
\end{center}
\end{figure}

\begin{figure}[!ht]
\begin{center}
\includegraphics[scale=0.75]{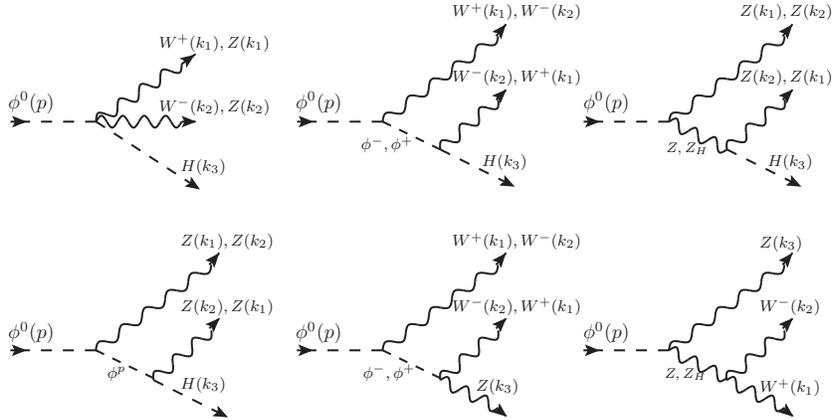}
\caption{Feynman diagrams representing tree-level contributions to $\Gamma_{\Phi^0}$.}
\label{feyn2ABC}
\end{center}
\end{figure}

The Feynman diagrams for two- and three-body decays contributing to $\Gamma_{\Phi^0}$ are shown in Figs.~\ref{feyn2VV} and \ref{feyn2ABC}, respectively. According to Fig.~\ref{feyn2VV}, for two-body decays we employ the vertices $\Phi^0 t\bar t, \Phi^0 ZZ, \Phi^0 HH$ given in Ref.~\cite{Han:2003wu}. Moreover, in the context of the linearized theory of LTHM the $\Phi^0 W_{\mu}^\pm\,W_{\nu}^\mp$ and $\Phi^0 {W_H}_{\mu}^\pm\,W_{\nu}^\mp$ vertexes vanish~\cite{Han:2003wu}.
Keeping in mind this facts, we now proceed to present the $\Phi^0 \to t\bar t$ decay width, which can be written as
\begin{equation}
\Gamma(\Phi^0 \to \bar t t )= \frac{N_c\,m_{\Phi^0} \, m_t^2}{64 \pi f^2} \left(1-4\frac{m_t^2}{m_{\Phi^0}^2} \right)^{3/2},
\end{equation}
where $m_{\Phi^0}$ is the heavy scalar mass, $N_c$ is the color factor equal to 3 for quarks and $f$ is the global symmetry breaking scale.

The $\Phi^0\to ZZ$ width decay is given as follows
\begin{eqnarray}
\Gamma(\Phi^0 \to ZZ)= \frac{g^4}{8^4\,\pi c_W^4}\frac{m_{\Phi^0}^3}{f^2}\frac{v^4}{m_Z^4}\left(1-4\frac{m_Z^2}{m_{\Phi^0}^2} \right)^{1/2}\left(1-4\frac{m_Z^2}{m_{\Phi^0}^2}+12\frac{m_Z^4}{m_{\Phi^0}^4} \right).
\end{eqnarray}

The $\Phi^0\to HH$ decay width is expressed as
\begin{equation}
\Gamma(\Phi^0 \to HH)=\frac{m_{\Phi^0}^3}{256 \pi f^2} \left(1-4\frac{m_H^2}{m_{\Phi^0}^2}\right)^{1/2}.
\end{equation}

Regarding to the three-body decays, these processes originate from a four-point vertex, as well as from Feynman diagrams mediated by neutral gauge bosons and heavy scalar bosons. Explicitly, the dominant three-body decays that contribute to $\Gamma_{\Phi^0}$ are $\Phi^0 \to ZZH$, $\Phi^0 \to WWZ$ and $\Phi^0 \to WWH$. The respective Feynman diagrams are depicted in Fig.~\ref{feyn2ABC}.

The decay width of the $\Phi^0$ scalar boson decaying to three bodies can be obtained by using the following equation
\begin{equation}
\frac{d\Gamma(\Phi^0 \to ABC)}{dx_a dx_b}=\frac{m_{\Phi^0}}{256\,\pi^3}|\mathcal{M}(\Phi^0 \to ABC)|^2,
\end{equation}
where $|\mathcal{M}|^2$ is the squared amplitude, $ABC$ represents the SM final states for each process presented in Fig.~\ref{feyn2ABC}. The amplitudes were calculated employing the Feynman rules given in Ref.~\cite{Han:2003wu}. The explicit form of these amplitudes is presented below
\begin{eqnarray}
\mathcal{M}(\Phi^0 \to WWH)=\frac{i\,g^2 v}{8 \sqrt{2} f}\Bigg[2g^{\mu\nu}+\frac{(k_1^\mu + 2k_3^\mu)(2k_1^\nu + k_2^\nu + 2k_3^\nu)}{(k_1+k_3)^2-m_{\Phi^0}^2}
+ \frac{(k_1^\mu + 2(k_2^\mu + k_3^\mu))(k_2^\nu + 2k_3^\nu)}{(k_2+k_3)^2-m_{\Phi^0}^2}\Bigg]\epsilon^*_\mu(k_1) \epsilon^*_\nu(k_2).
\end{eqnarray}
It must be recalled that $m_{\Phi^0}=m_{\Phi^+}=m_{\Phi^-}$.
\begin{eqnarray}
\mathcal{M}(\Phi^0 \to ZZH)&=&\frac{i g^2 v}{32 \sqrt{2} \,c_W^4 f} \Bigg[c_W^2 \Bigg(24 g^{\mu\nu} + 8\frac{(k_1^\mu + 2k_3^\mu)(2k_1^\nu + k_2^\nu + 2k_3^\nu)}{(k_1+k_3)^2-m_{\Phi^0}^2}
+ 8\frac{(k_1^\mu + 2(k_2^\mu + k_3^\mu))(k_2^\nu + 2k_3^\nu)}{(k_2+k_3)^2-m_{\Phi^0}^2}\Bigg)\nonumber\\ 
&+& 4g^2v^2\Bigg(\frac{g^{\mu\nu}-\frac{(k_{1}^\mu+k_{3}^\mu)(k_{1}^\nu+k_{3}^\nu)}{m_Z^2}}{(k_1+k_3)^2-m_Z^2}
+\frac{g^{\mu\nu}-\frac{(k_{2}^\mu+k_{3}^\mu)(k_{2}^\nu+k_{3}^\nu)}{m_Z^2}}{(k_2+k_3)^2-m_Z^2}\Bigg)\nonumber\\
&+&\frac{g^2(c^2-s^2)^2v^2}{c^2s^2}\Bigg(\frac{g^{\mu\nu}-\frac{(k_{1}^\mu+k_{3}^\mu)(k_{1}^\nu+k_{3}^\nu)}{m_{Z_H}^2}}{(k_1+k_3)^2-m_{Z_H}^2}
+\frac{g^{\mu\nu}-\frac{(k_{2}^\mu+k_{3}^\mu)(k_{2}^\nu+k_{3}^\nu)}{m_{Z_H}^2}}{(k_2+k_3)^2-m_{Z_H}^2}\Bigg)\Bigg]\epsilon^*_\mu(k_1) \epsilon^*_\nu(k_2),
\end{eqnarray}
\begin{eqnarray}
\mathcal{M}(\Phi^0 \to WWZ)&=& \frac{g^3v^2}{8\,\sqrt{2}\,c_W f^3}\Bigg[-if^2\Bigg(\frac{g^{\alpha\nu}(k_{1}^\mu+2(k_{2}^\mu+k_{3}^\mu))}{(k_2+k_3)^2-m_{\Phi^0}^2}
+\frac{g^{\alpha\mu}(2k_{1}^\nu+k_{2}^\nu+2k_{3}^\nu)}{(k_1+k_3)^2-m_{\Phi^0}^2}\Bigg)+\frac{2f^2g^{\alpha\beta}}{(k_1+k_2)^2-m_Z^2}\nonumber\\
&\times&\Bigg(\frac{(k_{1\,\beta}+k_{2\,\beta})(k_{1\,\gamma}+k_{2\gamma})}{m_Z^2}-g_{\beta\gamma}\Bigg)
\bigg((k_2^\gamma-k_1^\gamma)g^{\mu\nu}-(k_1^\mu + 2k_2^\mu)g^{\gamma\nu}+(2k_1^\nu+k_2^\nu)g^{\gamma\mu}\bigg)\nonumber\\
&-&\frac{(c^2-s^2)^2v^2g^{\alpha\beta}}{(k_1+k_2)^2-m_{Z_H}^2}
\Bigg(\frac{(k_{1\,\beta}+k_{2\,\beta})(k_{1\,\gamma}+k_{2\gamma})}{m_{Z_H}^2}-g_{\beta\gamma}\Bigg)
\bigg((k_2^\gamma-k_1^\gamma)g^{\mu\nu}-(k_1^\mu + 2k_2^\mu)g^{\gamma\nu}\nonumber\\
&+&(2k_1^\nu+k_2^\nu)g^{\gamma\mu}\bigg)\Bigg]\epsilon^*_\mu(k_1) \epsilon^*_\nu(k_2) \epsilon^*_\alpha(k_3).
\end{eqnarray}
To obtain the decay widths, we square the decay amplitudes with the aid of FeynCalc package~\cite{FeyC}. The integrations over three-body phase space were numerically performed.

\subsubsection{One-loop level $\Phi^0 \to WW$ decay}

\begin{figure}[!ht]
\begin{center}
\includegraphics[scale=0.55]{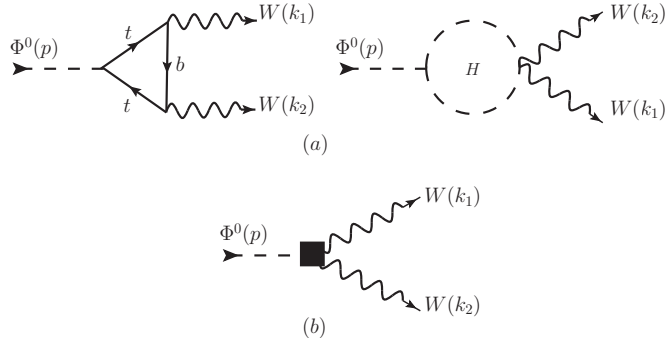}
\caption{Feynman diagrams contributing to the $\Phi^0 \to WW$ decay at one-loop level. (a) Dominant contributions. (b) Feynman diagram corresponding to counterterm contributions.}
\label{diaWW}
\end{center}
\end{figure}

Since the $\Phi^0 WW$ coupling is absent at the tree-level in the LTHM, it is interesting to study the one-loop level $\Phi^0 \to WW$ process. In the LTHM, this decay receives dominant contributions from SM top quark and the Higgs boson; when they circulate in the loops. The dominant Feynman diagrams can be appreciated in Fig.~\ref{diaWW}. After performing dimensional regularization for the one-loop amplitudes coming from Feynman diagrams in Fig.~\ref{diaWW}(a), we arrive to UV divergent contributions which cannot be analytically canceled, so it is necessary to implement the renormalization of the one-loop amplitudes in order to obtain finite contributions at the leading order. This is achieved by adding the corresponding counterterms to the ultraviolet infinite amplitude in $d=4$~\cite{Kniehl}. In Fig.~\ref{diaWW}(b), we present the Feynman diagram which represents counterterms that exactly cancel ultraviolet divergent contributions of the one-loop $\Phi^0WW$ vertex.

The amplitude for the $\Phi^0 \to WW$ decay can be written as
\begin{eqnarray}\label{PhiWW}
\mathcal{M}(\Phi^0 \to WW)&=&\frac{g^2 N_c V_{tb}^2 m_t^2 m_W^4}{16 \sqrt{2}\, \pi ^2 f m_{\Phi^0}^2 \left(m_{\Phi^0}^2-4 m_{W}^2\right)^2} \Big[A^{WW}(k_1^\mu k_1^\nu + k_2^\mu k_2^\nu)+ B^{WW}(k_1^\mu k_2^\nu + k_2^\mu k_1^\nu)\nonumber\\
&+& C^{WW} m_{\Phi^0}^2\,g^{\mu\nu}\Big]\epsilon^*_\mu(k_1)\epsilon^*_\nu(k_2),
\end{eqnarray}
where
\begin{eqnarray}
A^{WW}&=&4 \Bigg[m_W^2 C_0 + \Big(B_0(2) - B_0(3) - 2(1 + (m_b^2 + m_t^2) C_0 )  \Big)\nonumber\\
&+& \frac{(m_b^2 - m_t^2)}{m_W^2}\Big(B_0(2) + B_0(3) - 2 B_0(1)  + (m_b^2 - m_t^2) C_0 \Big)\Bigg]\nonumber\\
&+& 2\Bigg[ 1 - m_W^2 C_0 - 2 \left(B_0(2) - B_0(3) - m_b^2 C_0 \right)  + \frac{(m_b^2 -m_t^2)}{m_W^2}\nonumber\\
&\times&\Big(3 B_0(1) + B_0(3) - 4 B_0(2) + (m_b^2 - m_t^2)C_0\Big)\Bigg] \frac{m_{\Phi^0}^2} {m_W^2}\nonumber\\
&-& \Bigg[ \frac{(m_b^2 - m_t^2)}{m_{W}^2}(B_0(1) - B_0(2)) - m_W^2C_0 - (m_b^2 - m_t^2)C_0 \Bigg] \frac{m_{\Phi^0}^4}{m_W^4},
\end{eqnarray}
\begin{eqnarray}
B^{WW}&=&-\Bigg\{4 \Big(2 + 2 (m_b^2  +  m_t^2) C_0 - B_0(2)+B_0(3)\Big) - 4 m_W^2 C_0\nonumber\\
&-& 4\frac{(m_b^2-m_t^2)}{m_W^2}\Big(B_0(2) + B_0(3) - 2 B_0(1) + (m_b^2-m_t^2) C_0 \Big)\nonumber\\
&-& \Big(B_0(2) - B_0(3) + (m_t^2 - 3  m_b^2 - m_W^2) C_0 - 1\Big) \frac{m_{\Phi^0}^4}{m_W^4}\nonumber\\
&-& 2 \Bigg[\frac{(m_b^2-m_t^2)}{m_W^2} \Big(B_0(1)+B_0(2)-2 (B_0(3) + (m_b^2-m_t^2) C_0 )\Big)\nonumber\\
&-& \Big(B_0(2) - B_0(3) - 2(2 m_b^2 + m_t^2)C_0 - 3\Big) \Bigg] \frac{m_{\Phi^0}^2}{ m_W^2} \Bigg\},
\end{eqnarray}
\begin{eqnarray}
C^{WW}&=&\frac{(4 m_W^2 - m_{\Phi^0}^2)}{2 m_W^2}\Bigg[4 - \frac{m_{\Phi^0}^2}{m_W^2} + 2 (B_0(2)-B_0(3)) \frac{(m_b^2-m_t^2)}{m_W^2}\nonumber\\
&-& \left(\frac{m_b^2-m_t^2}{m_W^2}+1\right)(m_{\Phi^0}^2 - 2 (m_t^2 + m_W^2 - m_b^2)) C_0\Bigg].
\end{eqnarray}
Notice that the form factors $A^{WW}, B^{WW}$, and $C^{WW}$ are finite, being, $B_0(1)=B_0(0, m_b^2, m_t^2), B_0(2)=B_0(m_W^2, m_b^2, m_t^2), B_0(3)=B_0(m_{\Phi^0}^2, m_t^2, m_t^2)$ and $C_0=C_0(m_W^2, m_W^2, m_{\Phi^0}^2, m_t^2, m_b^2, m_t^2)$, the Passarino-Veltman scalar functions. Thus, the decay width of the $\Phi^0 \to WW$ process is
\begin{equation}\label{dwidthWW}
\Gamma(\Phi^0 \to WW)=\frac{\sqrt{m_{\Phi^0}^2-4m_W^2}}{16\,\pi m_{\Phi^0}^2}|\mathcal{M}(\Phi^0 \to WW)|^2.
\end{equation}
Let us mention that the contribution to the one-loop amplitude arising from the Higgs loop is exactly canceled by the respective counterterm.

\subsubsection{One-loop level $\Phi^0 \to \gamma V$ decays}

\begin{figure}[!ht]
\begin{center}
\includegraphics[scale=0.55]{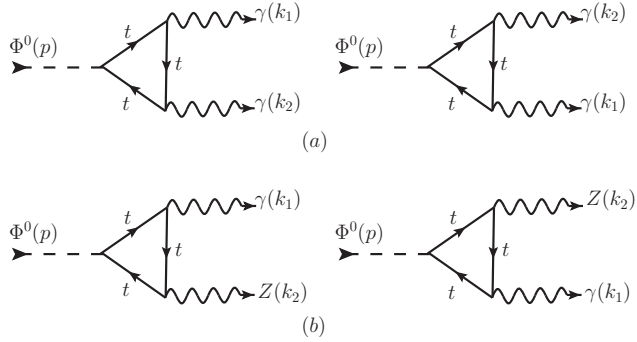}
\caption{Dominant Feynman diagrams contributing to the $\Phi^0 \to \gamma \gamma$ and $\Phi^0 \to \gamma Z$
decay at one-loop level.}
\label{feyn-diagrams-contrib}
\end{center}
\end{figure}

We will now describe the analytical expressions for the decay amplitudes of the $\Phi^0 \to \gamma V$ process, which were computed in the unitary gauge. In the LTHM these decays are only mediated by SM quarks and a new exotic top quark $T$. In specific, the heavy top quark circulating inside the fermionic-triangle loop for the $\Phi^0 \to \gamma Z$ process induces a non-zero contribution. However, in the amplitude, the contributions coming from this new exotic top quark are suppressed at least by two orders of magnitude with respect to the SM top quark contributions, which are the dominant ones between SM quarks. Therefore, we do not include this new exotic contribution in our calculations. In Fig. \ref{feyn-diagrams-contrib} we show the Feynman diagrams corresponding to $\Phi^0 \to \gamma \gamma$ and $\Phi^0 \to \gamma Z$ decays; inside the loops is circulating SM top quark. The contribution to $\Phi^0 \to \gamma \gamma$ with only fluctuating $T$ quark is null since there is no present the $Tt\gamma$ coupling.

The amplitude for the $\Phi^0 \to \gamma \gamma$ process is given by
\begin{eqnarray}\label{important1}
\mathcal{M}(\Phi^0 \to \gamma \gamma)=A^{\gamma \gamma} (k_1 \cdot k_2\, \,g^{\mu \nu} - k_1^\nu\,k_2^\mu)
\epsilon^*_\mu(k_1)\epsilon^*_\nu(k_2).
\end{eqnarray}
The coefficient $A^{\gamma \gamma}$ is expressed in terms of Passarino-Veltman (PV) scalar functions as it is observed below
\begin{eqnarray}
A^{\gamma \gamma}=\frac{g^2 N_c\, s_W^2 m_t^2}{9\,\sqrt{2}\,\pi^2\,f m_{\Phi^0}^2}((m_{\Phi^0}^2-4m_t^2)C_0(1)+2),
\end{eqnarray}
where $C_0(1)=C_0(m_\Phi^2, 0, 0, m_t^2, m_t^2, m_t^2)$ is the three-point PV scalar function. After performing some algebraic operations, the decay width of the $\Phi^0 \to \gamma \gamma$ process can be expressed as
\begin{eqnarray}
\Gamma(\Phi^0 \to \gamma \gamma)=\frac{|A^{\gamma \gamma}|^2\,m_{\Phi^0}^3}{64\,\pi}.
\end{eqnarray}

Let us study the $\Phi^0\to \gamma Z$ decay. The respective amplitude after some algebraic manipulations can be written as
\begin{eqnarray}\label{important2}
\mathcal{M}(\Phi^0 \to \gamma Z)&=&A^{\gamma Z} (k_1\cdot k_2\,g^{\mu \nu} - k_1^\nu k_2^\mu)\epsilon_\mu^*(k_1)\epsilon^*_\nu(k_2).
\end{eqnarray}
The coefficient $A^{\gamma Z}$ is explicitly given by
\begin{eqnarray}
A^{\gamma Z}= \frac{g^2 N_c s_W(3-8 s_W^2)m_t^2 m_Z^2}{36 \sqrt{2}\,\pi^2 c_W f (m_{\Phi^0}^2-m_Z^2)^2}
\Big((B_0(3)-B_0(4))+\frac{m_{\Phi^0}^2-m_Z^2}{2m_Z^2}(C_0(2)(4m_t^2+m_Z^2-m_{\Phi^0}^2)+2)\Big),
\end{eqnarray}
where $B_0(4)=B_0(m_Z^2,m_t^2,m_t^2)$ and $C_0(2)=C_0(m_{\Phi^0}^2,m_Z^2,0,m_t^2,m_t^2,m_t^2)$.
Finally, the decay width for this process can be read as follows
\begin{eqnarray}
\Gamma(\Phi^0 \to \gamma Z)= \frac{|A^{\gamma Z}|^2(m_{\Phi^0}^2-m_Z^2)^3}{32\,\pi\, m_{\Phi^0}^3}.
\end{eqnarray}
It should be stand out that the $A^{\gamma\gamma}$ and $A^{\gamma Z}$ coefficients are free of UV divergences and the Lorentz structures appearing in Eqs.~(\ref{important1}) and (\ref{important2}) satisfy gauge invariance. The numerical evaluation of the branching ratios for the $\Phi^0 \to \gamma V$ processes was carried out by means of LoopTools package~\cite{LoopTools}.

\section{Numerical results}\label{NUM-RES}
\begin{figure}[htb!]
\begin{center}
\includegraphics[scale=0.65]{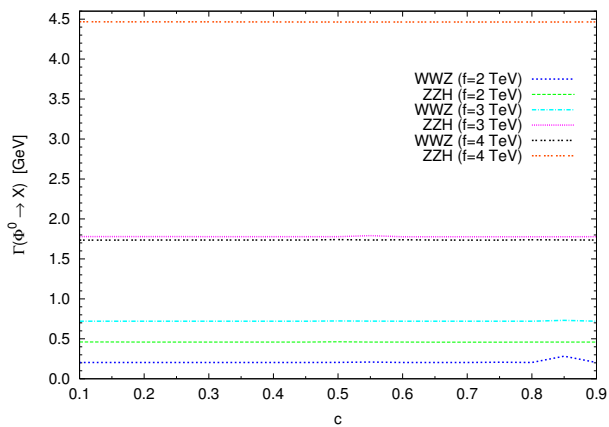}
\caption{Decay widths for the $\Phi^0\to WWZ$ and $\Phi^0\to ZZH$ processes as a function of the $c$ parameter for $f=2,3,4$ TeV.}
\label{widthc}
\end{center}
\end{figure}

\begin{figure}[htb!]
\centering
\subfigure[]{\includegraphics[scale=0.59]{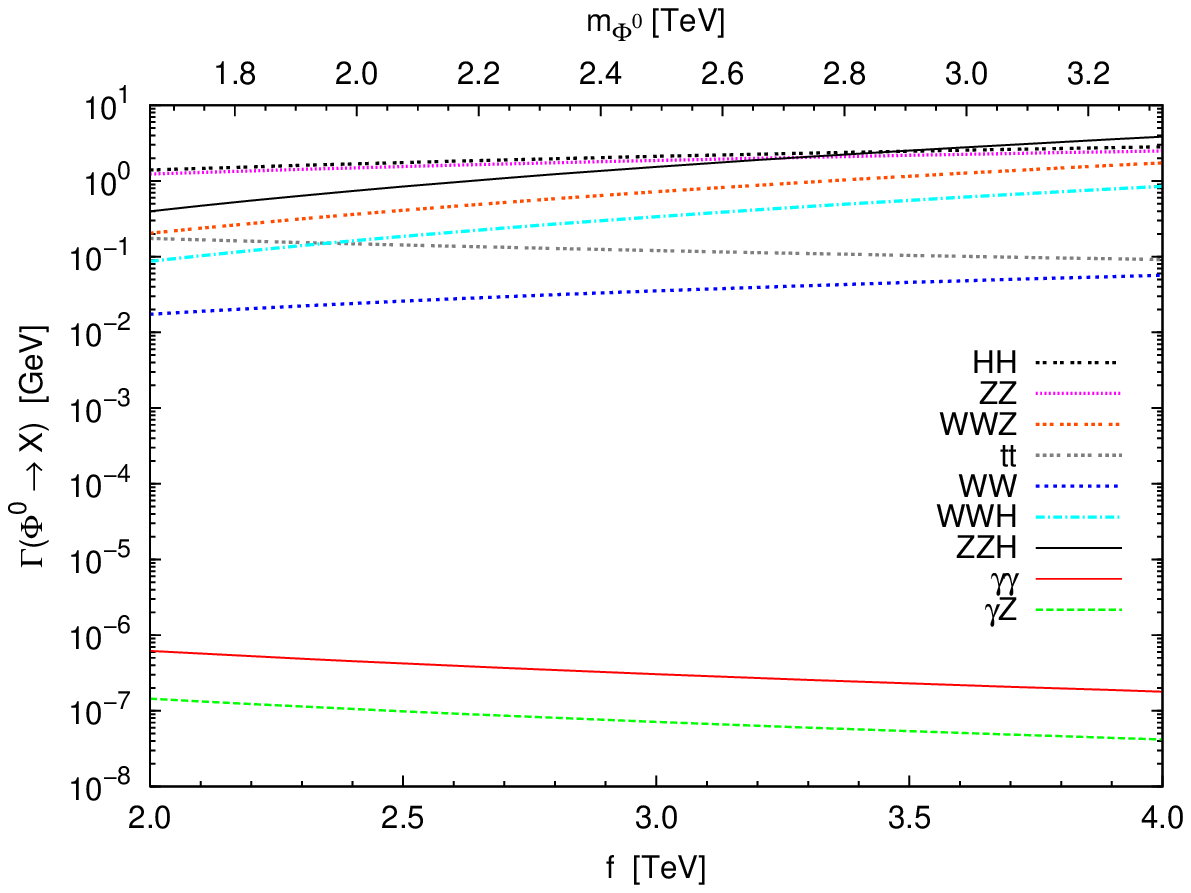}}\qquad
\subfigure[]{\includegraphics[scale=0.59]{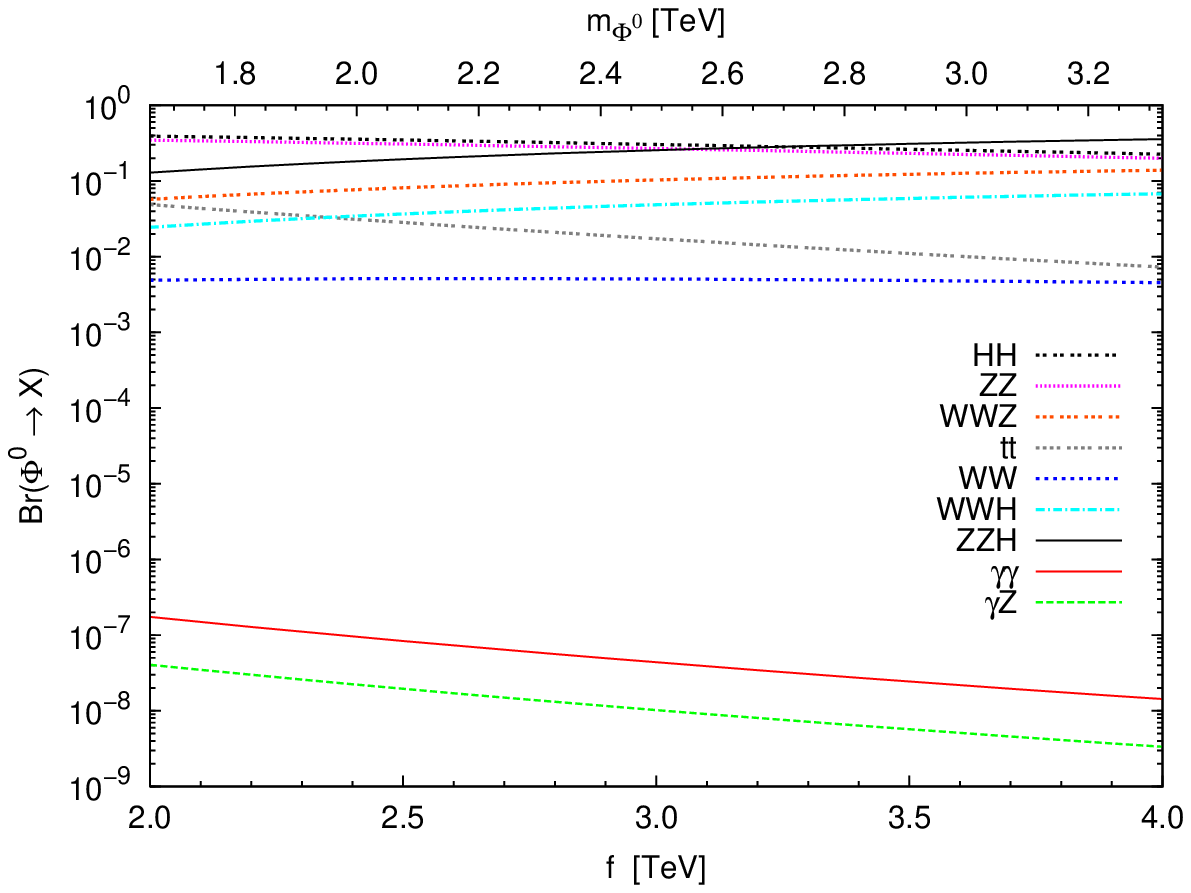}}
\caption{(a) Decay widths for the $\Phi^0\to X$ processes as a function of the $f$ parameter, where $X=HH, ZZ, WWZ, \bar{t}t, WW, WWH, ZZH, \gamma\gamma, \gamma Z$. (b) Branching ratios for the $\Phi^0 \to X$ processes depending on the $f$ parameter.}
\label{Brf}
\end{figure}

In this work we propose a scenario where $m_{\Phi^0}$ is around 1 TeV so that the one-loop level decays of the scalar into $WW, \gamma \gamma$ and $\gamma Z$ could help us to discern on the validity of the model or whether this model should be ruled out. In the plots that Fig.~\ref{widthc} shows, it is clearly observed that the decay widths of the $\Phi^0\to WWZ$ and $\Phi^0\to ZZH$ processes are not sensible to $c$ parameter variations. In Fig.~\ref{widthc}, a variation of the $c$ parameter between 0.1 and 0.9 is performed, for three distinct energy scales, i.e. $f=2$ TeV, $f=3$ TeV and $f=4$ TeV. Since only the $\Phi^0\to WWZ$ and $\Phi^0\to ZZH$ processes depend of $c$ and these are not the dominant decay modes of the $\Phi^0$ scalar particle, it is concluded that essentially the total decay width of the $\Phi^0$ boson will not depend of the $c$ parameter. Thus, we will assume a specific value of $c$ in such a way that the $Z_H$ contribution to $\Phi^0\to WWZ$ and $\Phi^0\to ZZH$ is magnified. This election seeks to avoid the decoupling in the $\Phi^0 Z_H Z$, $H Z_H Z$ and $Z_H WW$ interactions. Accordingly with Fig.~\ref{widthc}, it is assumed $c=0.85$, which corresponds to a small mixing angle and is consistent with a enhanced contribution of the $Z_H$ boson to $\Gamma(\Phi^0\to WWZ)$ and $\Gamma(\Phi^0\to ZZH)$. In Fig.~\ref{Brf}(a), the decay widths for the $\Phi^0\to HH, ZZ, tt, WWZ, WWH, ZZH, WW, \gamma\gamma, \gamma Z$ process are shown. In Fig.~\ref{Brf}(a), it can be observed that the dominant contributions come from two-body decays of $\Phi^0$ into $HH$ and $ZZ$, which are of the order of unities of GeV on the interval 2 TeV $< f <$ 4 TeV. The two-body decay widths of the $\Phi^0$ boson were calculated for the first time in Ref.~\cite{Twobwidhts}, however, these were not studied as a function of the energy scale $f$. We can observe that $\Gamma(\Phi^0\to WW)$ is of the order of $10^{-2}$ GeV over the interval 2 TeV $< f <$ 4 TeV. Conversely, the decay widths of the $\Phi^0\to \gamma\gamma$ and $\Phi^0\to \gamma Z$ processes are of the order of $10^{-7}$ GeV along the interval 2 TeV $< f <$ 2.3 TeV. Abreast, their respective branching ratios as a function of the energy scale $f$ are depicted in Fig.~\ref{Brf}(b). Before continuing with the analysis of results, we recall that $m_{\Phi^0}$ is a function of the energy scale $f$ at which the global symmetry is broken. Notice that the energy scale $f$ is the only free parameter with which we can play. The symmetry breaking scale is constrained by the experimental data being restricted to lower limits around 2-4 TeV~\cite{Reuter}. Since the mass of the $\Phi^0$ scalar boson is a monotonous increasing function in their dependence on the $f$ parameter, in our analysis we consider the range of study to be 2 TeV$<f<$4 TeV. Using this information, we obtain that the mass of the heavy scalar is ranging from $1.66$ TeV to $3.32$ TeV. The total decay width of the $\Phi^0$ boson includes the following decay modes: $HH, ZZ, \bar{t}t, WWZ, WWH, ZZH$.

From Fig.~\ref{Brf}(b), it is clearly seen that the $\gamma\gamma$ and $\gamma Z$ branching ratios are suppressed around 6 orders of magnitude compared to the dominant decay modes. In contrast, the associated branching ratio for the $\Phi^0\to WW$ process is of the order of $10^{-3}$ throughout the interval 2 TeV $< f <$ 4 TeV. Therefore, for these one-loop level processes, the decay channel corresponding to two SM $W$ bosons would be of great interest for experimental studies in production of hypothetical heavy neutral scalar bosons.

It is important to add that in the calculation of the total decay width of the $\Phi^0$ scalar boson it is found that the subdominant decays are: $\Phi^0 \to ZZH$, $\Phi^0 \to WWZ$, $\Phi^0 \to WWZ$ and $\Phi^0 \to t \bar t$, with decay widths being of the order of $10^{-1}$ GeV for $f$ between 2 TeV and 3 TeV.

Is relevant highlight that in the LTHM the lower limit for the $f$ parameter allows us to explore heavy neutral scalar masses greater than 1 TeV.

\subsection{Production of the heavy neutral scalar boson}
This section provides the results for the production cross section of the heavy scalar $\Phi^0$ in the context of the LTHM at LHC. To carry out this task, we used the version 2.1.1 of the WHIZARD package, which is a generic Monte-Carlo event generator for multi-particle processes at high-energy collisions~\cite{whizard}. This event generator was used in a previous work of us for the calculation of the associated production cross section of the $Z_H$ neutral gauge boson as a proof of the proper functioning of the WHIZARD package~\cite{Aranda}, where employed the CTEQ5 parton distribution function~\cite{{CTEQ6}}.
In this calculation we simulate proton-proton collisions at a center-of-mass energy of 14 TeV with the parton distribution function MSTW2008NLO~\cite{MSTW2008NLO}.
\begin{figure}[htb!]
\centering
\includegraphics[scale=0.85]{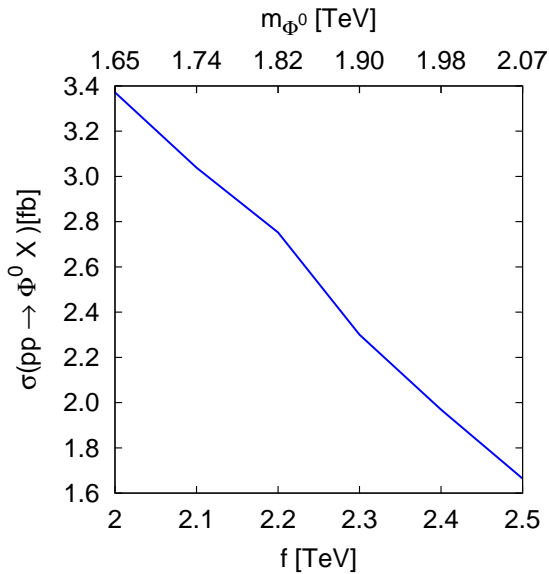}
\caption{Production cross section for $\Phi^0$ in proton-proton collisions at $\sqrt{s}= 14$ TeV
as a function of the $f$ parameter.
\label{phiprod}}
\end{figure}

In our simulations we calculate the cross secction as a function of the $f$ parameter in the range values between $2$ TeV and $2.5$ TeV, as it is shown in Fig.~\ref{phiprod}. The top axis represents the heavy scalar mass throughout the interval $1.66$ TeV$<m_{\Phi^0}<2.1$ TeV. In order to do this simulation we employed values from $f=2$ TeV until $f=2.5$ TeV. The behavior of the cross section is depicted in the Fig.~\ref{phiprod}, where we can see that the cross section is equal to $3.37$ fb for $f=2$ TeV and the corresponding value of the cross section for $f=2.5$ TeV is equal to $1.66$ TeV. As it was shown before, the branching ratios for the $\Phi^0\to\gamma\gamma, \gamma Z$ decays are very suppressed with respect to dominant decays of the $\Phi^0$ boson, therefore, these would not be of experimental interest. On the other hand, the branching ratio for the $\Phi^0\to WW$ process is of the order of $10^{-3}$ between 2 TeV $< f <$ 2.5 TeV, which results interesting because the LHC integrated luminosity at the final stage of operation is expected to be 3000 fb$^{-1}$ or maybe more~\cite{luminosity} which could reverse this plight. Continuing with the feasibility analysis of the LTHM, let us consider the smallest value studied for the mass of the scalar boson $\Phi^0$ which corresponds to the maximum value analyzed for the $\Phi^0$ production cross section $\sigma(pp\to \Phi^0 X)$. For these values, in the Gaussian approximation, it is found around unities of events for the $\Phi^0\to WW$ decay.

In this work is not intended to study in detail the mechanism of event production since we are doing an estimation exercise at best. Nevertheless, this analysis allows us to say that it would be difficult to observe a heavy neutral scalar resonance of the LTHM decaying into two $W$ bosons at the LHC. Thus, it would be of great interest to study the production of the $\Phi^0$ scalar boson in the context of the Compact Linear Collider since it would offers cleanest collisions with a considerably large integrated luminosity~\cite{CLIC}.

\section{Conclusions}\label{CONCLU}
The LTHM is constructed on the basis of a nonlinear sigma model with a $SU(5)$ global symmetry and the gauged subgroup $[SU(2)_1\otimes U(1)_1]\otimes[SU(2)_2\otimes U(1)_2]$, where the existence of new particles with masses of the order of TeV's is predicted, in particular, a new heavy neutral scalar particle known as $\Phi^0$. The $\Phi^0\to WW, \gamma V$ decays were proposed to explore the current parameter space of the LTHM model; it has given special emphasis to the decay of the $\Phi^0$ scalar boson into two $W$ bosons justified by the emergence of the recent experimental results previously discussed. Although the parameter space of the LTHM has been severely constrained, yet there is room to explore the $\Phi^0$ decay into two $W$ bosons as a function of the energy scale $f$. We have studied physical regions accordingly to the experimental bounds and phenomenological results; in specific, we have proposed the following range of study for the $f$ parameter: 2 TeV$<f<$4 TeV. Thus, we find out that the $\Phi^0$ mass is between 1.66 TeV and 3.32 TeV and the branching ratio for the $\Phi^0 \to WW$ decay is of the order of $10^{-3}$ on this energy interval. Regarding to the $\Phi^0 \to \gamma \gamma$ and $\Phi^0 \to \gamma Z$ processes, their branching ratios at best take values of $10^{-7}$ and $10^{-8}$, respectively. We also simulate the production cross section of the heavy scalar particle in proton-proton collision at $\sqrt{s}=14$ TeV, being found that for $f=2$ TeV could be estimated around one ten of events for the $\Phi^0 \to WW$ decay at LHC in the best case and in ideal conditions.

\section*{Acknowledgments}
This work has been partially supported by CONACYT, SNI-CONACYT, and CIC-UMSNH.\\


\end{document}